\documentclass[5p]{elsarticle}

\usepackage{amsmath}
\usepackage{amssymb}
\usepackage[colorlinks=true,allcolors=blue]{hyperref}
\usepackage{mathtools}
\usepackage{tikz}
\usepackage{booktabs}
\usetikzlibrary{shapes.geometric,angles,quotes,calc}

\newcommand\rossby\varepsilon

\newcommand\vecb[1]{\mathbf{#1}}
\newcommand\diff{\mathrm{d}}
\newcommand\I{\mathrm{i}}

\newcommand\bfvec[1]{\mathbf{#1}}

\begin{document}
    
\title{SQG Point Vortex Dynamics with Order Rossby Corrections}

\author[1]{Mac Lee}
\ead{mal004@ucsd.edu}
\author[2,3]{Stefan G.\ Llewellyn Smith}
\ead{sgls@ucsd.edu}

\affiliation[1]{Department of Physics, University of California, San Diego CA
92093-0319, USA}
\affiliation[2]{Department of Mechanical and Aerospace Engineering, Jacobs
School of Engineering, University of California, San Diego CA 92093-0411, USA}
\affiliation[3]{Scripps Institution of Oceanography, University of California,
San Diego CA 92093-0209, USA}

\begin{abstract}
  Quasi-geostrophic flow is an asymptotic theory for flows in rotating systems
  that are in geostrophic balance to leading order. It is characterized by the
  conservation of (quasi-geostrophic) potential vorticity and weak vertical
  flows. Surface quasigeostrophy (SQG) is the special case when the flow is
  driven by temperature anomalies at a horizontal boundary. The next-order
  correction to QG, QG+, takes into account ageostrophic effects. We investigate
  point vortx dynamics in SQG+, building on the work of Weiss~\cite{weiss2022}.
  The conservation laws for SQG point vortices that parallel the 2D Euler case
  no longer exist when ageostrophic effects are included. The trajectories of
  point vortices are obtained explicitly for the general two-vortex case in SQG
  and SQG+. For the three-vortex case, exact solutions are found for rigidly
  rotating and stationary equilibria consisting of regular polygons and
  collinear configurations. As in the 2D case, only certain collinear
  vortex configurations are rigid equilibria. Trajectories of passive tracers
  advected by point vortex systems are studied numerically, in particular their
  vertical excursions, which are non-zero because of ageostrophic effects.
  Surface trajectories can manifest local divergence even though the underlying
  fluid equations are incompressible.
\end{abstract}


\maketitle

\section{Introduction}

Vortices play an important role in plasma physics, geophysics and planetary
sciences. They are observed in the Earth's atmosphere and oceans and also on
other planets, e.g.~the cyclones at Jupiter's poles that form vortex
crystals~\citep{adriani2018,siegelman2022b,siegelman2022a}. Vortices are also
observed in the laboratory, including in two-dimensional magnetised electron
columns~\citep{fine1995,jin1998,schecter1999,jin2000}. In the far field, intense
vortices (local disturbances in the vorticity field) can be approximated by
point vortices~\citep{newton2001}.  In fact, numerical calculations using point
vortices can quite accurately capture the dynamics of a more complex vorticity
configuration under certain circumstances~\citep{marchioro1983}. This appealing
reduction of a complex PDE system to a discrete system has inspired numerous
theoretical studies, from the vortex atom~\citep{thomson1867} to
fundamental research on the mathematical underpinning of these objects for more
complex geometries~\citep{aref1979,aref2002,aref2009} to
spheres~\citep{bogomolov1977,bogomolov1979}, and more general curved
surfaces~\citep{hally1980,kimura1999}. Two-dimensional plane point vortex
dynamics has been extensively studied~\citep{newton2001}. In particular
equilibrium solutions in rigid body motion include triangles, vortex polygons,
and collinear vortices, while the general motion of more than three vortices is
not integrable.

Strong rotation and stratification in large-scale geophysical flows leads to
quasi-two-dimensional motion characterized by geostrophic and hydrostatic
equilibrium at leading order. The resulting asymptotic description of the motion
is known as quasigeostrophy (QG) \cite{vallis2017}. Potential vorticity is the
key dynamic quantity, with its inversion giving the velocity
field~\cite{mcintyre2000}. As a result, singularities in (potential) vorticity have been
investigated as dynamic entities~\cite{gryanik2000}. Surface quasigeostrophy
(SQG) is the special case in which the potential vorticity in the flow interior
is zero, so that the flow is due to temperature anomalies on the
boundary~\citep{held1995,lapeyre2006,lapeyre2017}. The relevant small quantity
in QG is the Rossby number $\rossby = U/fL$. When it increases, corresponding to
stronger flows or small scales, QG becomes less accurate and the ageostrophic or
unbalanced part of the flow can no longer be neglected.

Muraki et al.~\cite{muraki1999} derived the next order correction terms to the
QG equations, while Hakim et al.~\cite{hakim2002} found an ansatz that
simplifies the Poisson equations in this formulation. Using Muraki's
formulation, Weiss \cite{weiss2022} studied point vortices under QG with
$O(\rossby)$ correction terms. Weiss's study focussed mainly on special cases of
two and three point vortices and did not obtain general solutions. Motivated by
understanding 3D effects in models of advection, Taylor et
al.~\cite{taylor2015,taylor2016} studied SQG point vortices, but did not use a
fully consistent expression for the $O(\rossby)$ velocity field.

This paper studies the equations of motion and conservation laws for point
vortices in SQG and SQG+, obtaining full solutions for the two-vortex case and
for three-vortex equilibria including dipoles, triangles, collinear point
vortices, and polygons. The trajectories of passive tracers advected by such
configurations are also examined. In \S\,\ref{sec:sqg}, we introduce the
equations of motion in SQG+. In \S\,\ref{sec:conservation_laws}, we study the
conservation laws in SQG and SQG+. In \S\,\ref{sec:two-pv}, we study the
dynamics of systems of two point vortices and derive the trajectories of such
objects. In \S\,\ref{sec:three-pv}, we analyse systems of three point vortices.
We further generalize the results to $N$ point vortices in \S\,\ref{sec:n-pv}.
In \S\,\ref{sec:vertical_motion}, we study the vertical motion as a result of
the order Rossby correction. \S\ref{sec:conlusions} concludes the study. Some
technical results are relegated to the Appendices.

\section{SQG with order Rossby corrections}\label{sec:sqg}

We study the dynamics of collections of point vortices. Following the
derivation in \cite{weiss2022} for QG point vortices, we write the total
potential temperature field as a collection of such singularities, with $\theta$
representing the potential temperature field and $\Gamma_i$ the circulation
strength of the $i$-th vortex located at $\vecb{x}_i$,
\begin{equation}
  \theta(\vecb{x}) = \sum_{i=1}^N \Gamma_i \delta(\vecb{x}-\vecb{x}_i(t)).
\end{equation}
In Weiss's notation, the governing equations for point vortices in SQG+ are
\begin{subequations}
  \label{eqs:velocity_equations}
  \begin{multline}
    \vecb{u}(\vecb{x}_i) = \sum_{j=1}^N \left(\Gamma_j\vecb{u}_0(\vecb{x}_i-\vecb{x}_j) + \rossby \Gamma_j^2 \vecb{u}_1^s(\vecb{x}_i - \vecb{x}_j) \right) \\
    + \rossby \sum_{\substack{j=1\\k=j+1}}^N \Gamma_j\Gamma_k\vecb{u}_1^p (\vecb{x}_i-\vecb{x}_j,\vecb{x}_i-\vecb{x}_k),
  \end{multline}
  where
  \begin{equation}
    \vecb{u}_0(\vecb{x}) = \frac{1}{2\pi|\vecb{x}|^3} \begin{pmatrix}
      -y \\ x \\ 0
    \end{pmatrix},
  \end{equation}
  \begin{equation}
    \vecb{u}_1^s(\vecb{x}) = \frac{x^2+y^2-8z^2}{4\pi^2|\vecb{x}|^8} \begin{pmatrix}
      -y \\ x \\ 0
    \end{pmatrix},
  \end{equation}
  \begin{equation}
    \vecb{u}_1^p(\vecb{x}_1, \vecb{x}_2) = \frac{\tilde{\vecb{u}}}{4\pi^2|\vecb{x}_1|^5|\vecb{x}_2|^5},
  \end{equation}
  and
  \begin{equation}
    \vecb{\tilde{u}} = (\tilde{u}_1, \tilde{u}_2, \tilde{u}_3),
  \end{equation}
  using the notation
  \begin{multline}
    \tilde{u}_1 = 3|\vecb{x}_1|^2(y_1z_2^2+2y^2z_1z_2) \\
    + 3|\vecb{x}_2|^2(y_2z_1^2+2y_1z_1z_2) - |\vecb{x}_1|^2|\vecb{x}_2|^2(y_1+y_2)
  \end{multline}
  \begin{multline}
    \tilde{u}_2 = -3|\vecb{x}_1|^2(x_1z_2^2+2x^2z_1z_2) \\
    - 3|\vecb{x}_2|^2(x_2z_1^2+2x_1z_1z_2) + |\vecb{x}_1|^2|\vecb{x}_2|^2(x_1+x_2)
  \end{multline}
  \begin{equation}
    \tilde{u}_3 = 3(x_2y_1-x_1y_2)(|\vecb{x}_2|^2z_1-|\vecb{x}_1|^2z_2).
  \end{equation}
\end{subequations}
The vertical velocity of tracers away from the surface is not zero, but is given by
\begin{multline}
  \label{eq:w}
  w = 3\rossby \sum_{\substack{j=1\\k=j+1}}^N
  \frac{\Gamma_j\Gamma_k}{4\pi^2|\vecb{x}_i-\vecb{x}_j|^5|\vecb{x}_i-\vecb{x}_k|^5} \\
  \times \left\{(x_i-x_k)(y_i-y_j)-(x_i-x_j)(y_i-y_k)\right\}\\
  \times\left\{|\vecb{x}_i-\vecb{x}_k|^2(z_i-z_j)-|\vecb{x}_i-\vecb{x}_j|^2(z_i-z_k)\right\}.
\end{multline}
However, this quantity is zero on the surface. For clarity, we can write the
surface equations of motion in complex notation. In this notation, the equation
of motion of the point vortex at $z_i$ is given by
\begin{equation}\label{eq:complex_number_pv_2d_velocity}
  \dot{z}_i = -\sum_{j \neq i}^{N}\frac{\I\Gamma_j(z_j-z_i)}{2\pi\left|z_i-z_j\right|^3},
\end{equation}
which is identical to the expression for point vortices in a two-dimensional
plane but with an inverse cube law. With $O(\rossby)$ corrections, this equation
becomes
\begin{multline}
  \label{eq:z_time_evo}
  \dot{z}_i = \sum_{j\neq i}^N \frac{\I\Gamma_j(z_i-z_j)}{2\pi|z_i-z_j|^3}\left(1+\frac{\rossby\Gamma_j}{2\pi|z_i-z_j|^3}\right) \\
  + \sum_{j\neq i}^N \sum_{k\neq i,j}^N\frac{\I\rossby\Gamma_j\Gamma_k(2z_i-z_j-z_k)}{8\pi^2|z_i-z_j|^3|z_i-z_k|^3}.
\end{multline}

\section{Hamiltonians and conservation laws}\label{sec:conservation_laws}

\begin{figure}
  \centerline{\includegraphics[width=2.5in]{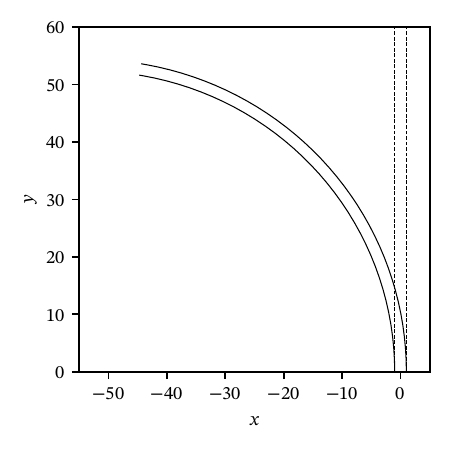}}
  \caption{SQG (dashed) and SQG+ (solid) dipole trajectories for $0 \le t \le
  600$. For the SQG+ trajectory, the Rossby number $\rossby$ is $0.3$. The
  initial position of the point vortex pairs is $(\pm1, 0, 0)$ and both vortex
  pairs have circulations of $\pm\pi$.}
  \label{fig:dipole}
\end{figure}

\begin{figure*}[t]
  \includegraphics[width=\textwidth]{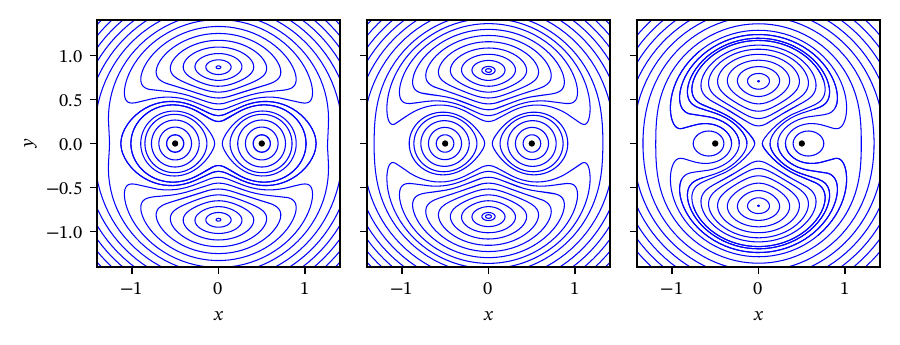}

  \includegraphics[width=\textwidth]{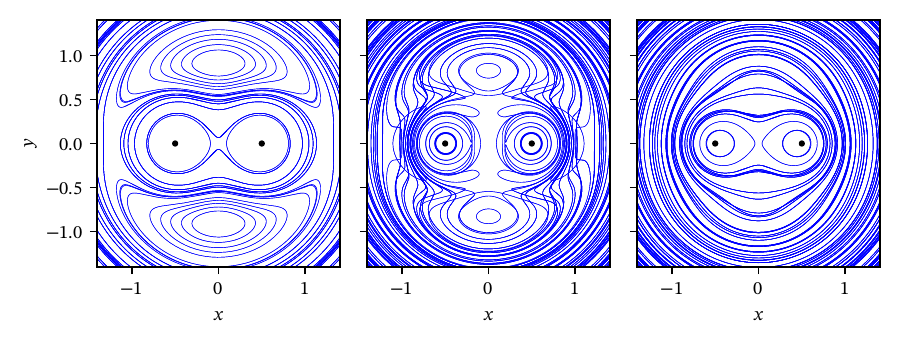}
  \caption{Trajectories of passive tracers in the co-rotating frame advected by
  two point vortices at $x=\pm0.5$ with identical circulations. The first row
  shows trajectories with zero Rossby number, where the second row shows
  trajectories with $\rossby=0.3$. The three plots, from left to right, show the
  trajectories of point vortices initially at $z=0$, $-0.25$, and $-0.5$. The
  plots in the second and third columns are the horizontal positions of the
  passive tracers.}
  \label{fig:two-pv-trajectory}
\end{figure*}

The Hamiltonian structure of point vortex dynamics has been studied for general
$\alpha$ models~\cite{badin2018}. In an $\alpha$ model, the relationship between
an active tracer $\zeta$ and the streamfunction $\psi$ is given by
\begin{equation}
  -\zeta = (-\Delta)^{\alpha/2} \psi.
\end{equation}
Both 2D Euler and SQG dynamics fall into this class of flows: for 2D Euler
systems, $\alpha$ is $2$; for SQG, $\alpha$ is $1$. The Hamiltonian for SQG is
given by
\begin{equation}
  H = -\sum_i^N\sum_{j\neq i}^N \frac{\Gamma_i\Gamma_j}{2\pi\left|z_j-z_i\right|}.
\end{equation}

The generalised position and momentum of each individual point vortex is then
given by
\begin{subequations}
  \begin{equation}
    Q_i = \Gamma_i x_i,
  \end{equation}
  \begin{equation}
    P_i = \Gamma_i y_i.
  \end{equation}
\end{subequations}
The conservation of the Hamiltonian can be confirmed by showing
\begin{equation}
  \frac{\diff H}{\diff t} = [H, H]_P = 0,
\end{equation}
where
\begin{equation}
  [f, g]_P =
  \sum_{i} \left(\frac{\partial f}{\partial P_i}\frac{\partial g}{\partial Q_i} 
  - \frac{\partial g}{\partial P_i}\frac{\partial f}{\partial Q_i}\right)
\end{equation}
is the Poisson bracket with respect to $z_i$ and $P_i$.

The first and second moments of the system, i.e.~the total linear and angular
momenta, are given by
\begin{equation}
  Q = \sum_i^N Q_i \qquad P = \sum_i^N P_i \qquad I = \sum_i^N\Gamma_i |z_i|^2.
\end{equation}
These quantities are conserved. It follows that a stationary centre of vorticity
can be defined as
\begin{equation}\label{eq:z_vc}
  z_{\text{vc}} = \frac{\sum_i\Gamma_iz_i}{\sum_i\Gamma_i}
\end{equation}
when $\sum_i\Gamma_i\neq 0$.

With order Rossby corrections, none of the above conservation laws remain. We
are not aware of the existence of a Hamiltonian for an SQG+ point vortex system.
The total linear momentum of an $N$-vortex system under SQG+ is no longer
conserved, but evolves in time according to
\begin{equation}
  \frac{d}{dt}\sum_i^N\Gamma_iz_i = \sum_i^N \sum_{j=i+1}^N\frac{\I\rossby\Gamma_i\Gamma_j(\Gamma_j-\Gamma_i)(z_i-z_j)}{4\pi^2\left|z_i-z_j\right|^6}.
\end{equation}
Since the centre of vorticity, defined in \eqref{eq:z_vc}, is proportional to
the linear momentum, it is not stationary. The total angular momentum is also not
conserved. Its time evolution under SQG+ is given by
\begin{equation}
  \frac{d}{dt} \sum_i^N\Gamma_i\left|z_i\right|^2 = \sum_i^N \sum_{j=i+1}^N \frac{\I\rossby\Gamma_i\Gamma_j(\Gamma_i+\Gamma_j)(z_iz_j^* - z_i^*z_j)}{2\pi^2\left|z_i-z_j\right|^6}.
\end{equation}
While there are no conservation laws for the general, $N$-vortex system in SQG+,
the conservation of total linear momentum and total angular momentum is restored
if every point vortex in the system has identical circulation. Moreover, for the
special case of a two vortex system, total angular momentum is always conserved,
regardless of circulation strengths of each individual point vortex.

\section{Trajectories of two point vortices}\label{sec:two-pv}

\begin{figure*}[t]
  \includegraphics[width=0.49\textwidth]{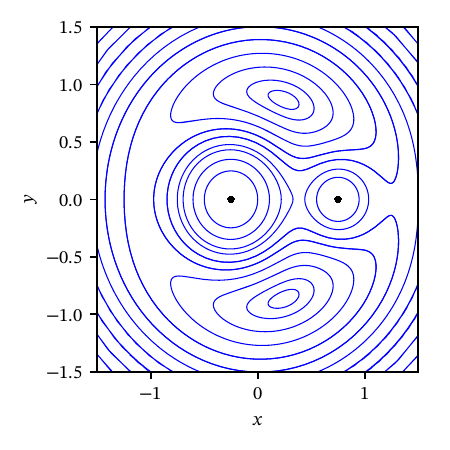}
  \includegraphics[width=0.49\textwidth]{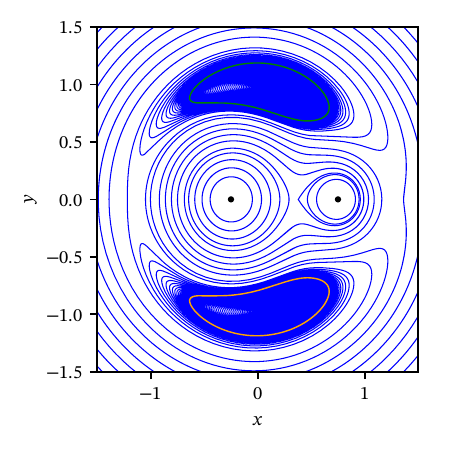}
  \caption{Projection of the trajectories of passive tracers in the co-rotating
  frame advected by two point vortices onto a horizontal plane. The circulations
  of the point vortices are, from left to right, $3\pi$ and $\pi$. Both plots
  are at the surface, $\rossby=0$ on the left, and $0.3$ on the right.}
  \label{fig:two-pv-unequal-circulation-trajectory}
\end{figure*}

\begin{figure*}[t]
  \begin{center}
    \includegraphics[width=\textwidth]{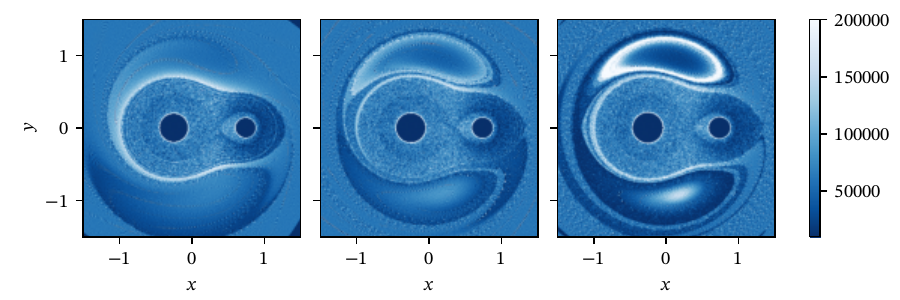}
  \end{center}
  \caption{Passive tracer density at different times. Initially, the passive
  tracers were uniformly distributed with $62500$ tracers per unit area over
  $(\pm2, \pm2, 0)$. The parameters are otherwise identical to that of
  Fig.~\ref{fig:analytical_divergence_unequal_pv}. From left to right, $t=1.5$,
  $5$, $15$. The speed of a passive tracer scales inversely with the cube of the
  distance to a point vortex.}
  \label{fig:particle_density}
\end{figure*}

\subsection{Analytical solution in SQG}

In a system of two point vortices, from
\eqref{eq:complex_number_pv_2d_velocity}, the velocity of each point vortex is
always orthogonal to the direction of the other point vortex. It then follows
that
\begin{equation}
  \frac{d}{dt}\rho^2 = \left[\rho^2, H\right]_P = 0,
\end{equation}
where $\rho = \left|z_1 - z_2\right|$. The separation between point vortices in
a two vortex system is hence conserved. The distance of any point vortex from
the centre of vorticity is also conserved, since
\begin{equation}
  \label{eq:rho1_rho}
  \rho_i = \left|z_i-z_{\text{vc}}\right| = \frac{\Gamma_j}{\Gamma_i + \Gamma_j} \rho,
\end{equation}
where $i,j\in \{1,2\}$. Using this property, $z_i$ can be written as
\begin{equation}
  \dot{z_i} = \I\rho_i\omega \exp\I\phi,
\end{equation}
where $\phi=\arg(z_i-z_j)$ is the relative phase between the two point vortices,
and $\omega=\dot{\phi}$ is the rate of rotation. Therefore,
\begin{equation}
  \label{eq:z1_time_evo_fixed_rho}
  \I\rho_i\omega \exp\I\phi
  = \frac{\I\Gamma_j\left(z_i-z_j\right)}{2\pi\rho^3}
  = \frac{\I\Gamma_j\exp\I\phi}{2\pi\rho^2}.
\end{equation}
Combining \eqref{eq:rho1_rho} and \eqref{eq:z1_time_evo_fixed_rho}, the relative
phase between two point vortices changes at a rate
\begin{equation}\label{eq:two_vortex_qg_rotation_rate}
  \phi(t) = \frac{\Gamma_1+\Gamma_2}{2\pi\rho^3}t + \phi_0.
\end{equation}
\begin{subequations}
  When $\Gamma_1+\Gamma_2\neq 0$, the trajectory of a point vortex in a system
  of two point vortices is
  \begin{equation}
    z_1(t) = z_1(0) +
      \frac{\Gamma_2\rho \exp\I\phi_0}{\Gamma_1 + \Gamma_2}\left\{\exp\left(\I\frac{\Gamma_1 + \Gamma_2}{2\pi\rho^3}t\right) - 1\right\}.
  \end{equation}
  Since the centre of vorticity is stationary and remains collinear with the two
  point vortices, the two point vortices revolve around the centre of vorticity
  at rate $\omega$.

  This analysis breaks down when $\Gamma_1+\Gamma_2=0$. In this special case,
  the point vortices travel in a straight line with
  \begin{equation}
    \label{eq:z1_trajectory}
    z_1(t) = z_1(0) +
      \frac{\I\Gamma_2 \exp\I\phi_0}{2\pi\rho^2} t.
  \end{equation}
\end{subequations}
The straight line is orthogonal to the direction of their separation and the
constant speed is $\Gamma/2\pi\rho^2$ with $\Gamma_1=-\Gamma_2=\Gamma$. It is
informative to cast \eqref{eq:z1_trajectory} in terms of the centre of
vorticity,
\begin{equation}
  z_1(t) = z_{\text{vc}} + \frac{\Gamma_2\rho}{\Gamma_1+\Gamma_2}\exp \left\{\I\left(\frac{\Gamma_1+\Gamma_2}{2\pi\rho^3}t + \phi_0\right)\right\}.
\end{equation}
In the limiting case of $\Gamma_1+\Gamma_2=0$, the distance of the point
vortices to the centre of vorticity approaches infinity.

\subsection{Analytical solution in SQG+}

The equation of motion of point vortex 1 in SQG+ is
\begin{equation}
  \label{eq:z1_time_evo}
  \dot{z}_1 =
  \frac{\I\Gamma_2\left(z_1-z_2\right)}{2\pi\rho^3}\left(1+\frac{\rossby\Gamma_2}{2\pi\rho^3}\right).
\end{equation}
Since the total linear momentum is not conserved under SQG+, the centre of
vorticity is no longer stationary, except for the special case
$\Gamma_1=\Gamma_2$. With two point vortices, the  distance between the point
vortices is conserved under time evolution even with ageostrophic effects:
\begin{align}
  \frac{d}{dt}\rho^2 =
  \sum_{i\neq j}^2
    \left(1+\frac{\rossby\Gamma_i}{2\pi\rho^3}\right)\frac{\I\Gamma_i\left|z_j-z_i\right|^2}{2\pi\rho^3} + c.c.
  = 0
\end{align}
We find
\begin{equation}
    \frac{d}{dt}(z_1-z_2) = \I\omega \rho \exp\I\phi
    = \I\left(\frac{\Gamma_1+\Gamma_2}{2\pi\rho^2} + \rossby\frac{\Gamma_1^2+\Gamma_2^2}{4\pi^2\rho^5}\right)\exp{\I\phi}.
\end{equation}
This leads to
\begin{equation}
  \label{eq:omega_2pv}
  \omega = \frac{\Gamma_1+\Gamma_2}{2\pi\rho^3} + \rossby\frac{\Gamma_1^2+\Gamma_2^2}{4\pi^2\rho^6}.
\end{equation}
\begin{subequations}
Hence when $\omega\neq0$, the trajectory of point vortex 1 is
  \begin{equation}
    z_1(t) = z_1(0) +
      \frac{\Gamma_2\exp{\I\phi_0}}{2\pi\omega\rho^2}\left(\frac{\rossby\Gamma_2}{2\pi\rho^3}+1\right)\left(\exp{\I\omega t}-1\right)
  \end{equation}
  When $\omega=0$, the system moves in a straight line following
  \begin{equation}
    z_1(t) = z_1(0)+
      \frac{\I\Gamma_2\exp{\I\phi_0}}{2\pi\rho^2}\left(1+\frac{\rossby\Gamma_2}{2\pi\rho^3}\right)t.
  \end{equation}
\end{subequations}
By definition, the centre of vorticity is always collinear with the two point
vortices. Since it is not stationary in SQG+, it must be revolving around some
common centre alongside the two point vortices. The equation of motion of the
centre of vorticity is
\begin{equation}
  \frac{d}{dt}z_{\text{vc}} = \frac{\Gamma_1\dot{z}_1 + \Gamma_2\dot{z}_2}{\Gamma_1+\Gamma_2}
  = -\frac{\I\rossby\Gamma_1\Gamma_2(\Gamma_1-\Gamma_2)\exp{\I\phi}}{4\pi^2(\Gamma_1+\Gamma_2)\rho^5}.
\end{equation}
When $\omega\neq0$,
\begin{multline}
  z_\text{vc} = z_\text{vc}\big|_{t=0} - \frac{\rossby\Gamma_1\Gamma_2(\Gamma_1-\Gamma_2)\rho \exp{\I\phi_0}}{2\pi\rho^3(\Gamma_1+\Gamma_1)^2+\rossby(\Gamma_1+\Gamma_2)(\Gamma_1^2+\Gamma_2^2)} \\
    \times\left\{\exp\left(\I\left[\frac{\Gamma_1+\Gamma_2}{2\pi\rho^3}+\rossby\frac{\Gamma_1^2+\Gamma_2^2}{4\pi^2\rho^6}\right]t\right)-1\right\};
\end{multline}
when $\omega=0$,
\begin{equation}
  z_\text{vc}(t) = z_\text{vc}(0) - \frac{\I\rossby\Gamma_1\Gamma_2(\Gamma_1-\Gamma_2)\exp{\I\phi_0}}{4\pi^2(\Gamma_1+\Gamma_2)\rho^5}t.
\end{equation}
In other words, the centre of vorticity revolves around the point
\begin{equation}
  z_\text{vc}(t) +
    \frac{\rossby\Gamma_1\Gamma_2(\Gamma_1-\Gamma_2)\rho \exp{\I\phi_0}}{2\pi\rho^3(\Gamma_1+\Gamma_1)^2+\rossby(\Gamma_1+\Gamma_2)(\Gamma_1^2+\Gamma_2^2)},
\end{equation}
which moves to infinity when $\Gamma_1+\Gamma_2=0$. Since the centre of
vorticity is collinear with the two point vortices, the two point vortices 
also revolve around this point.
In the limit $\rossby\rightarrow0$, this reduces to
\begin{align}
  \begin{split}
    z_c
    = \frac12(z_1+z_2)\big|_{t=0} + \frac12\frac{\Gamma_1-\Gamma_2}{\Gamma_1+\Gamma_2}\rho \exp{\I\phi_0}
    = \frac{\Gamma_1z_1+\Gamma_2z_2}{\Gamma_1+\Gamma_2},
  \end{split}
\end{align}
which is the centre of vorticity. The centre of vorticity is stationary when
$\rossby=0$.
In SQG+, the trajectories of the two point vortices trace out concentric circles
about $z_c$ even when the total circulation is zero, as the rate of rotation
$\omega$ given by \eqref{eq:omega_2pv} is not zero. The trajectories become
straight lines when $\omega$ is zero, corresponding to
\begin{equation}
  \label{eq:zero-rotation}
  \Gamma_1 + \Gamma_2 + \rossby\frac{\Gamma_1^2 + \Gamma_2^2}{2\pi\rho^3} = 0.
\end{equation}
Solving for $\Gamma_1$, we obtain
\begin{align}
  \label{eq:zero-rotation-order-rossby-condition}
  \begin{split}
    \Gamma_1 &=
    -\frac{\pi\rho^3}{\rossby} \pm \frac{\pi\rho^3}{\rossby} \left\{
      1 - \frac{\rossby\Gamma_2}{\pi\rho^3}
      - \frac{\rossby^2\Gamma_2^2}{2\pi^2\rho^6}
      \right\}
    + O(\rossby^2) \\
    &\approx -\Gamma_2 \left(1- \frac{\rossby\Gamma_2}{2\pi\rho^3}\right).
  \end{split}
\end{align}
From \eqref{eq:zero-rotation}, the rotation rate is zero in SQG when the
circulations are equal and opposite. In the presence of ageostrophic effects,
the rotation rate is non-zero even when the total circulation is zero. This
motion is shown in Fig.~\ref{fig:dipole}, where the trajectories of a pair of
point vortices are plotted for both SQG and SQG+.

\subsection{Trajectories of passive tracers}

By definition, SQG+ point vortices reside on the surface. According to
\eqref{eq:w}, they do not have vertical movement. However, they advect passive
tracers in their vicinity, and studying the trajectories and the vertical
movement of these passive tracers resulting from the ageostrophic correction in
SQG+ sheds light into the effects of the next-order correction on vertical
transport. In both SQG and SQG+, two point vortices revolve about some
stationary point. Since the point vortices revolve around a common, stationary
centre, all our plots are made in the co-rotating frame of the point vortices in
which they appear stationary. Similar studies for point vortices in QG+
were previously carried out in \cite{weiss2022}.

The trajectories of passive tracers advected by two identical point vortices is
shown in Figure~\ref{fig:two-pv-trajectory}. With two point vortices of equal
circulation, the passive tracer trajectories are all periodic. At the surface,
the passive tracer trajectories of SQG and SQG+ are qualitatively similar. At
depth, their behaviours start to diverge.
The topology of the trajectories changes as depth increases. At the surface, the
area around the two vortices can be divided into four regions. In the region in
the immediate vicinity of the vortices, the passive tracers revolve around the
point vortices as they would around a single point vortex. Further away, there
is a narrow band where the passive tracers orbit both point vortices. With the
two point vortices lined up along the $x$-axis, the regions above and below
along the $y$-axis form their own closed loops, with passive tracers moving
about a point that forms an equilateral triangle with the two point vortices.
Further out, the passive tracers revolve around the entire two point vortex
system. These outer trajectories become more circular further away they are from
the point vortices. As depth increases, the innermost region shrinks. In SQG,
the thin band surrounding both point vortices is pinched until it no longer
wraps around both vortices, at which point the two sets of isolated orbits above
and below the system connect, forming a dumbbell shaped region. The change in
behaviour with respect to depth seems to be more rapid in SQG+ than its SQG
counterpart.

\begin{figure}
  \begin{center}
    \includegraphics[width=3in]{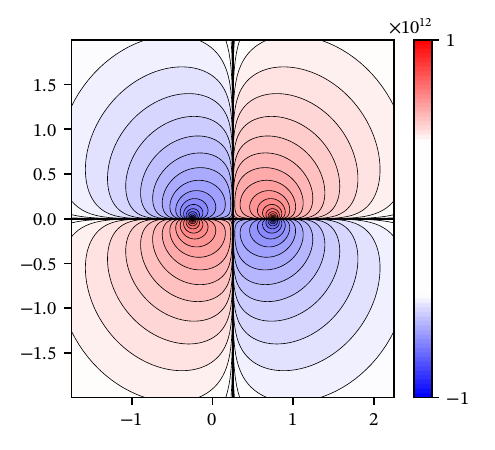}
  \end{center}
  \caption{The divergence of a system of two point vortices. The point vortex on
  the left has circulation $\pi$ and the one on the right has circulation
  $3\pi$. The vortices are situated at $(-0.25, 0, 0)$ and $(0.75, 0, 0)$.}
  \label{fig:analytical_divergence_unequal_pv}
\end{figure}

For two point vortices with different circulations, there are more fundamental
differences between SQG+ and SQG passive tracer trajectories. Trajectory
plots at the surface for SQG shows trajectories that are similar to the previous
set of plots, with the only visible difference being the breaking of left-right
mirror symmetry. Compared to the  plot of the equal strength case, these
trajectories are skewed, and the lobes above and below the point vortex system
are shifted to the right. All the trajectories are still periodic. On the other
hand, the SQG+ trajectory plot shows regions where the trajectories spiral
inward or outward. In the right panel of
Figure~\ref{fig:two-pv-unequal-circulation-trajectory}, the green curves
highlights the ``attractor'' of the system, where all tracers in the upper lobe
and all tracers outside of the orange curve in the lower lobe spiral toward. All
tracers in the lower lobe spiral away from the closed orange curve. This is
because the horizontal divergence of the velocity field at the surface is
non-zero when we account for order Rossby corrections. The horizontal divergence
of the velocity field is
\begin{equation}
  \label{eq:divergence_field}
  \Xi = \frac{3i\rossby\Gamma_1\Gamma_2 \left\{\overline{\Delta}_1\Delta_2-\Delta_1\overline{\Delta}_2\right\}}{8\pi^2\left|\Delta_1\right|^3\left|\Delta_2\right|^3}\left\{\frac{1}{\left|\Delta_1\right|^2}-\frac{1}{\left|\Delta_2\right|^2}\right\},
\end{equation}
where $\Delta_i \equiv z - z_i$. This divergence field as visualised in
Figure~\ref{fig:analytical_divergence_unequal_pv} is positive in quadrants I and
III and negative in quadrants II and IV. This explains the spiralling
trajectories for the fluid parcels around two point vortices with different
circulations. The change in passive tracer density due to the non-zero horizontal
divergence can be seen in Fig.~\ref{fig:particle_density}. This suggests that
unlike its cousin in SQG, a system of two point vortices with unequal
circulation strengths in SQG+ does not have a streamfunction.

\begin{figure*}[h]
  \includegraphics[width=\textwidth]{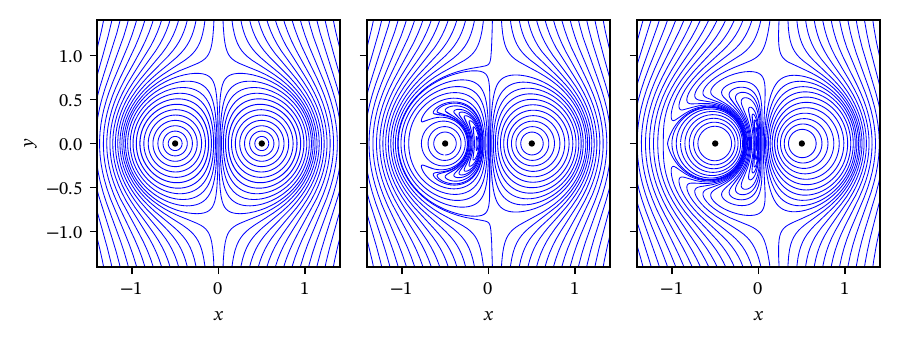}

  \includegraphics[width=\textwidth]{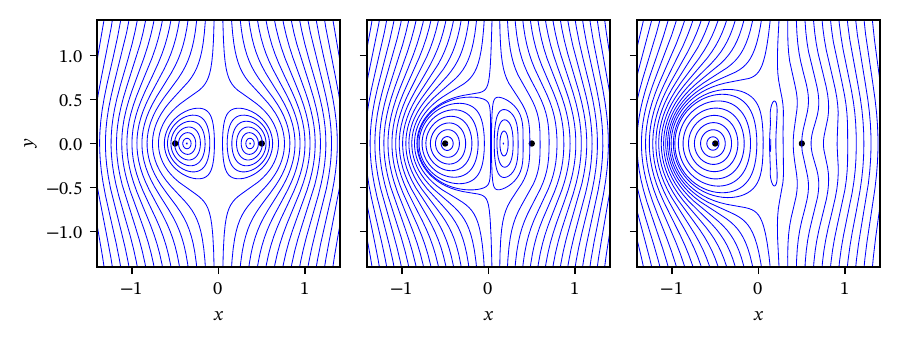}
  \caption{Projection of the trajectories of passive tracers onto a horizontal
  plane in a frame of the two moving vortices. First row: trajectories of
  passive tracers at the surface, second row: trajectories at $z=-0.7$. The
  Rossby numbers for the plots from left to right are $0$, $0.1$, and $0.3$. The
  strengths of the two vortices are equal and opposite in the first column. For
  the second column, the circulations of the vortices are $\pi$ and
  $-3.4928402508$. The circulations of the vortices in the third column are
  $\pi$ and $-4.16414263055$. For the second and third columns, the circulations
  of the point vortices are chosen such that $\omega=0$ using
  \eqref{eq:zero-rotation-order-rossby-condition}.}
  \label{fig:two-pv-zero-omega-trajectories}
\end{figure*}

When $\omega$ is zero, two point vortices travel in a straight line. In the
frame where the two point vortices are stationary as in
Figure~\ref{fig:two-pv-zero-omega-trajectories}, the mirror symmetry that is
present in the passive tracer trajectories across the $y$-axis is eliminated
when $O(\rossby)$ effects are included. A small pocket of isolated trajectories
develops between the vortex with the strength smaller in magnitude and the
$y$-axis. This pocket grows in size as the Rossby number is increased.

In short, while the point vortex trajectories of pairs of SQG+ point vortices
only slightly differ from that of an identical set of SQG point vortices, the
trajectories of passive tracer advected by them are not divergence-free at the
surface, unlike that of the latter. A streamfunction does not exist for such
surface motions.

\section{Trajectories of three point vortices}\label{sec:three-pv}

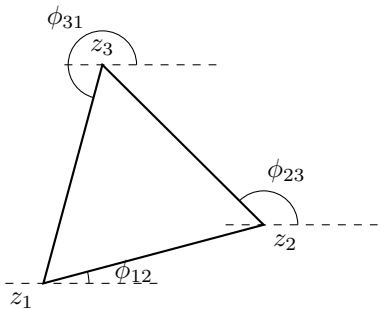
\begin{figure}
  \begin{center}
    \begin{tikzpicture}
      \def\side{3} 

      \coordinate (z1) at (0, 0); 
      \coordinate (z2) at (\side, 0); 
      \coordinate (z3) at ({\side/2}, {sqrt(3)*\side/2}); 

      \coordinate (rz2) at ({\side*cos(15)}, {\side*sin(15)}); 
      \coordinate (rz3) at ({\side/2*cos(15) - (sqrt(3)*\side/2)*sin(15)},
                            {\side/2*sin(15) + (sqrt(3)*\side/2)*cos(15)}); 

      \coordinate (h1) at ($(z1) + (1,0)$); 
      \coordinate (h2) at ($(rz2) + (1,0)$); 
      \coordinate (h3) at ($(rz3) + (1,0)$); 

      \draw[thick] (z1) -- (rz2) -- (rz3) -- cycle;

      \node[below left] at (z1) {$z_1$};
      \node[below right] at (rz2) {$z_2$};
      \node[above] at (rz3) {$z_3$};

      \draw[dashed] ($(z1) + (-0.5,0)$) -- ($(z1) + (1.5,0)$); 
      \draw[dashed] ($(rz2) + (-0.5,0)$) -- ($(rz2) + (1.5,0)$); 
      \draw[dashed] ($(rz3) + (-0.5,0)$) -- ($(rz3) + (1.5,0)$); 

      \pic [draw=black, angle radius=0.6cm, "$\phi_{12}$", angle eccentricity=2] {angle=h1--z1--rz2};

      \pic [draw=black, angle radius=0.45cm, "$\phi_{23}$", angle eccentricity=1.75] {angle=h2--rz2--rz3};

      \pic [draw=black, angle radius=0.45cm, "$\phi_{31}$", angle eccentricity=1.75] {angle=h3--rz3--z1};
    \end{tikzpicture}
  \end{center}
  \caption{A triangle and its angles.}
  \label{fig:triangle-illustration}
\end{figure}
The equation of motion for a point vortex in a system of three point vortices
is, for $i,j,k\in\{1,2,3\}$,
\begin{multline}
  \label{eq:three_pv_single_pv_time_evo}
  \dot{z}_i =
  \frac{\I\Gamma_j(z_i-z_j)}{2\pi|z_i-z_j|^3}\left(1+\frac{\rossby\Gamma_j}{2\pi|z_i-z_j|^3}\right) \\
  + \frac{\I\Gamma_k(z_i-z_k)}{2\pi|z_i-z_k|^3}\left(1+\frac{\rossby\Gamma_k}{2\pi|z_i-z_k|^3}\right) \\
  + \frac{\I\rossby\Gamma_j\Gamma_k(2z_i-z_j-z_k)}{4\pi^2|z_i-z_j|^3|z_i-z_k|^3}.
\end{multline}
As explained in \S\ref{sec:conservation_laws}, in general, the total linear and
angular momenta are not conserved in the presence of ageostrophic effects. The
distance between any two point vortex is not conserved either, but evolves in
time according to
\begin{multline}
  \label{eq:rho2_time_evo_angles}
  \frac{d}{dt}\rho_{12}^2 =
  \frac{\I\Gamma_3 \rho_{12} \exp{\I(\phi_{13}-\phi_{12})}}{2\pi\rho_{13}^2}
  \left\{1 + \frac{\rossby}{2\pi}\left(\frac{\Gamma_2}{\rho_{12}^3} + \frac{\Gamma_3}{\rho_{13}^3}\right)\right\} \\
  - \frac{\I\Gamma_3 \rho_{12} \exp{\I(\phi_{23}-\phi_{12})}}{2\pi\rho_{23}^2}
  \left\{1 + \frac{\rossby}{2\pi}\left(\frac{\Gamma_1}{\rho_{12}^3} + \frac{\Gamma_3}{\rho_{23}^3}\right)\right\} \\
  + c.c..
\end{multline}
In a triangle with vertices at $z_1$, $z_2$, and $z_3$, $\phi_{13} - \phi_{12} =
\angle z_1$ and $\phi_{23} - \phi_{12} = \pi - \angle z_2$, modulo $2\pi$, as
illustrated in Fig.~\ref{fig:triangle-illustration}, so that the complex coefficients of
the right-hand side of \eqref{eq:rho2_time_evo_angles} can  be written as
\begin{subequations}
  \begin{align}
    \I\exp{\I(\phi_{13} - \phi_{12})} + c.c. &= -2\sin \angle z_1  \\
    \I\exp{\I(\phi_{23} - \phi_{12})} + c.c. &= -2\sin \angle z_2.
  \end{align}
\end{subequations}
Since
\begin{equation}
  \rho_{12}\sin\angle z_1 = \frac{2A}{\rho_{13}}
\end{equation}
where $A$ is the area bounded by the triangle formed by $z_1$, $z_2$ and
$z_3$, we can simplify the above into
\begin{multline}
  \label{eq:three_pv_rho_time_evo}
  \frac{d}{dt}\rho_{12}^2 = \frac{2A}{\pi}\Bigg\{\frac{\Gamma_3}{\rho_{23}^3}\left[1 + \frac{\rossby}{2\pi}\left(\frac{\Gamma_1}{\rho_{12}^3} + \frac{\Gamma_3}{\rho_{23}^3}\right)\right] \\
  - \frac{\Gamma_3}{\rho_{13}^3}\left[1 + \frac{\rossby}{2\pi}\left(\frac{\Gamma_2}{\rho_{12}^3} + \frac{\Gamma_3}{\rho_{13}^3}\right)\right]
  \Bigg\}.
\end{multline}
With $\rossby=0$, this reduces to (68) of \cite{badin2018}, which is the SQG
analogue of (8) in~\cite{aref1979}. The absence of a length conservation law in
general means that we cannot as easily solve for the trajectories of the point
vortices in a general way as was done in the preceding sections on two point
vortices. However, there are special cases in which length is conserved.

\subsection{Equilateral triangle in SQG}

According to \eqref{eq:three_pv_rho_time_evo}, at $\rossby=0$, the distance
between point vortices is conserved if they are all equidistant from one
another. In this configuration, the centre of vorticity is stationary, provided
that $\Gamma_1 + \Gamma_2 + \Gamma_3 \neq 0$. The triangular configuration of
these point vortices then rotates around the centre of vorticity, where the rate
of rotation is  obtained using \eqref{eq:three_pv_single_pv_time_evo} as
\begin{equation}
  \omega = \frac{\Gamma_1 + \Gamma_2 + \Gamma_3}{2\pi\rho^3},
\end{equation}
where $\rho$ is the separation between the point vortices. When the total
circulation is zero, the system does not rotate at all. Instead, the system
translates at a velocity
\begin{equation}
  v = \frac{\sqrt{2 (\Gamma_1^2 + \Gamma_2^2 + \Gamma_3^3)}}{4\pi\rho^2}.
\end{equation}

\begin{figure*}[t]
  \includegraphics[width=\textwidth]{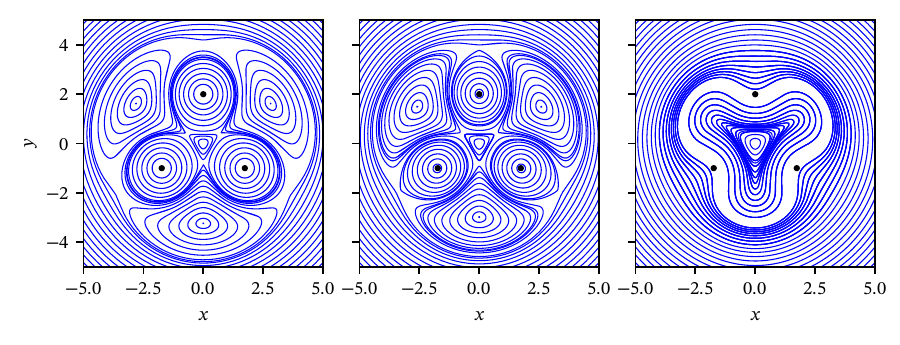}

  \includegraphics[width=\textwidth]{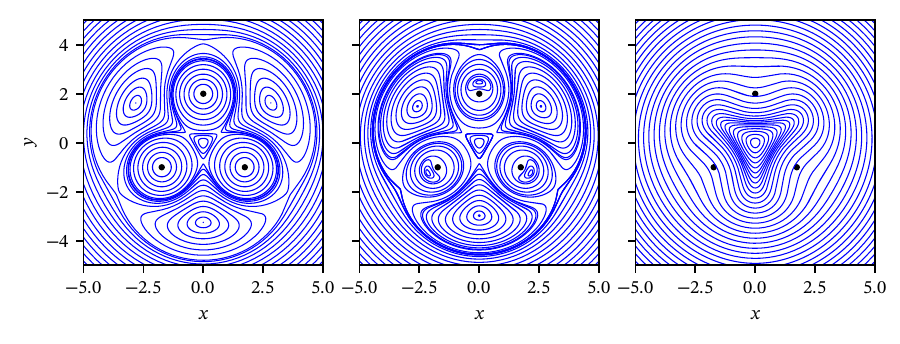}
  \caption{Trajectories of passive tracers in the co-rotating frame in an
  equilateral triangle configuration with the point vortices of identical
  circulation of $\pi$. The plots from left to right show tracers were
  initially at $z=0$, $-1$, and $-2$. The black dots are the vortices. The
  Rossby number $\rossby$ is $0$ and $0.3$ for the top and bottom rows.}
  \label{fig:equilateral_triangle_trajectories}
\end{figure*}
 
\subsection{Equilateral triangle in SQG+}

With $O(\rossby)$ corrections, the centre of vorticity is no longer stationary,
and distances between point vortices are no longer conserved for arbitrary
circulations. However, in the special case of $\Gamma_1 = \Gamma_2 = \Gamma_3
\equiv \Gamma$, we are able to recover key results similar to those of SQG. For
a system with three point vortices of identical strength arranged in an
equilateral triangle, the centre of vorticity is located at the centre of the
triangle and is stationary. The rate of rotation of the system about this centre
is
\begin{equation}
  \omega = \frac{3\Gamma}{2\pi\rho^3} \left(1 + \frac{\rossby\Gamma}{\pi\rho^3}\right).
\end{equation}

The trajectories of passive tracers advected by an equilateral triangle at two
different depths are plotted in
Figure~\ref{fig:equilateral_triangle_trajectories}. In the outer parts of the
system, the passive tracers revolve around the three point vortices as if they
were a single point vortex at the centre. Closer to the system, the passive
tracers either follow centres of revolution outside the triangular point vortex
configuration or they revolve around the three vortices. At the very centre of
the system, there is a region in which passive tracers revolve around the centre
of the system in a rounded triangular orbit.

\subsection{Collinear point vortices in SQG and SQG+}

\begin{figure*}[t]
  \begin{center}
    \includegraphics[width=0.49\textwidth]{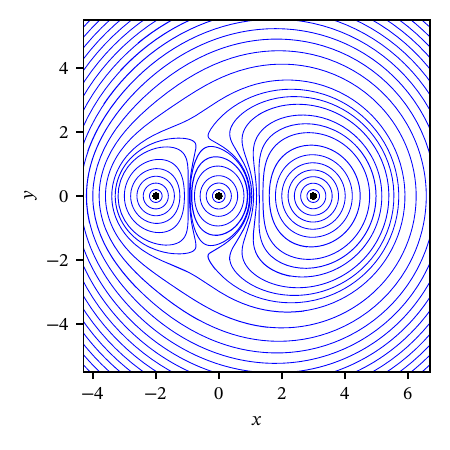}
    \includegraphics[width=0.49\textwidth]{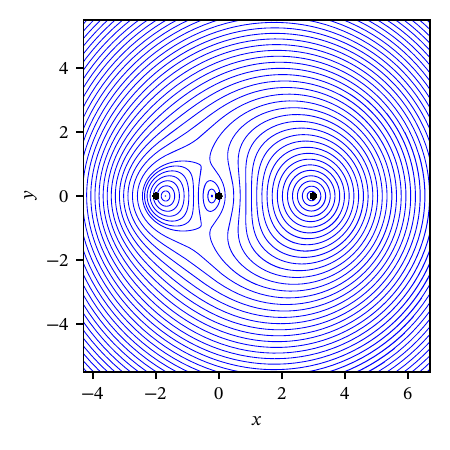}
  \end{center}
  \caption{Trajectories of passive tracers advected by a set of three collinear
  point vortices in the co-rotating frame. The three point vortices, from left
  to right, have circulations of $2126\pi/1053$, $-2\pi$, and $3\pi$. The ratio
  $R$ is $2/3$. The trajectories in the figure on the left are at the surface,
  whereas those on the right are at a depth of $-1.5$.}
  \label{fig:collinear_3pv_trajectories}
\end{figure*}

The result \eqref{eq:three_pv_rho_time_evo} suggests that three collinear
point vortices should move like a rigid body, but this is not the case in 2D as
noted by \cite{aref1979}.  For three instantaneously collinear vortices in SQG,
if the separation between any two vortices stays constant, the rotation rate of
the line connecting vortices $z_1$ and $z_2$ is
\begin{equation}
  \label{eq:collinear_omega_12}
  \omega_{12} = \frac{1}{2\pi\rho_{12}}\left\{
    \frac{\Gamma_1}{\rho_{12}^2} + \frac{\Gamma_2}{\rho_{12}^2}
    + \frac{\Gamma_3}{(\rho_{12} + \rho_{23})^2} - \frac{\Gamma_3}{\rho_{23}^2}
  \right\}.
\end{equation}
Similarly, the rotation rate of the line connecting vortices $z_2$ and
$z_3$ is
\begin{equation}
  \label{eq:collinear_omega_23}
  \omega_{23} = \frac{1}{2\pi\rho_{23}}\left\{
    \frac{\Gamma_1}{(\rho_{12} + \rho_{23})^2} - \frac{\Gamma_1}{\rho_{12}^2}
    + \frac{\Gamma_2}{\rho_{23}^2} + \frac{\Gamma_3}{\rho_{23}^2}
  \right\},
\end{equation}
which is clearly not the same as $\omega_{12}$. This shows that, in general,
three collinear vortices in SQG do not move together as a rigid body. We can
find the class of configurations where the three point vortices do rotate as a
rigid body by imposing the requirement that
$\omega_{12}=\omega_{23}\equiv\omega$. Rewriting $\rho_{12} / \rho_{23}$ as $R$,
we obtain a quintic equation that describes the necessary conditions for rigid
body motion,
\begin{multline}
  (\Gamma_2 + \Gamma_3) R^5 + (2\Gamma_2 + 3\Gamma_3)R^4 + (\Gamma_2 + 3\Gamma_3)R^3
  - (3\Gamma_1 + \Gamma_2)R^2 \\
  - (3\Gamma_1+2\Gamma_2)R - (\Gamma_1+\Gamma_2)=0.
\end{multline}
This equation is the SQG analogue of (37) in \cite{aref2009}. There is no
general analytical solution to this equation. One special case is worth
mentioning. With two identical vortices, $\Gamma_1=\Gamma_3\equiv\Gamma$, $R$ is
identically $1$. This result is obvious from symmetry arguments . Using
\eqref{eq:collinear_omega_12}, the rate of rotation of this configuration is
\begin{equation}
  \omega = \frac{\Gamma+4\Gamma_c}{8\pi\rho^3}.
\end{equation}
A plot of the trajectories of a chosen stable configuration is shown in
Figure~\ref{fig:collinear_3pv_trajectories}. For SQG+, imposing the collinear
requirement on \eqref{eq:collinear_omega_12} and \eqref{eq:collinear_omega_23}
yields an eleventh order equation,
\begin{multline}\label{eq:collinear_criteria_sqg_plus}
  AR^{11} + BR^{10} + CR^{9} + DR^{8} + ER^{7} + FR^{6} \\
  + GR^{5} + HR^{4} + IR^{3} + JR^{2} + KR + L = 0.
\end{multline}
The coefficients are given in \ref{sec:collinear-criteria}. This is the SQG+ analogue to (37) in \cite{aref2009}. There
is no analytical solution to this equation.


\section{Trajectories of vortex polygons with more than three point
vortices}\label{sec:n-pv}

\begin{figure*}[t]
    \centering
    \includegraphics[width=\textwidth]{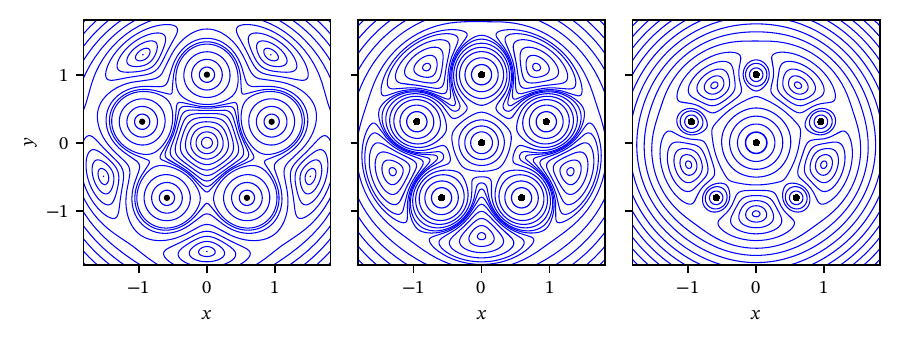}
    \caption{Passive tracer trajectories induced by systems of vortices arranged in a regular pentagon. The Rossby number $\rossby$ is zero. The outer vortices all have a circulation strength of $\pi$. The strength of the vortex in the centre is, from left to right respectively, $0$, $\pi$ and $10\pi$.}
    \label{fig:pentagon-corotate-ro0}
\end{figure*}

\begin{figure*}
    \centering
    \includegraphics[width=0.48\textwidth]{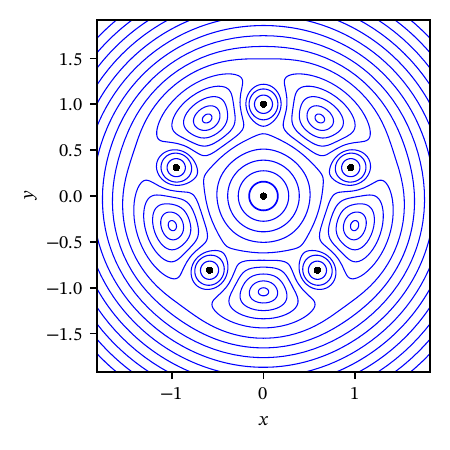}
    \includegraphics[width=0.48\textwidth]{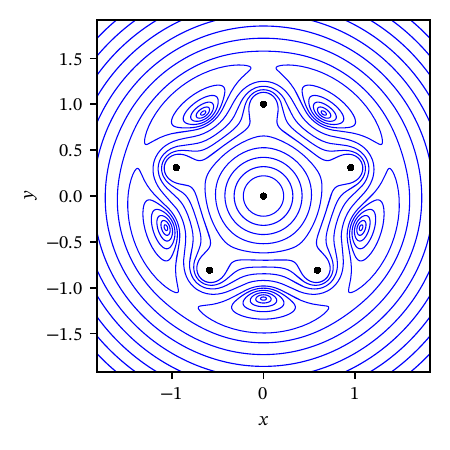}
    \caption{Passive tracer trajectories induced by systems of vortices arranged
    in a regular pentagon. The strength of the vortices in the pentagon is
    $\pi$, and the strength of the central vortex is $10\pi$. The Rossby number
    is $0$ on the left and $0.3$ on the right.}
    \label{fig:pentagon-ro0.3}
\end{figure*}

\begin{figure*}
    \begin{center}
        \includegraphics[width=\textwidth]{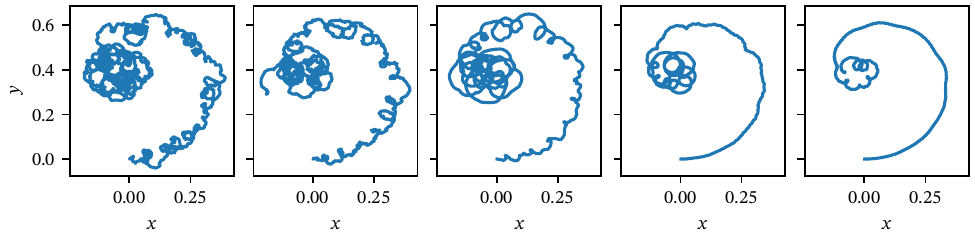}
    \end{center}
    \caption{The trajectories of centres of mass of an evenly spaced grid
    of $128\times128$ passive tracers initially distributed over $(\pm5, \pm5, z)$.
    From left to right, the passive tracers are initially
    placed at $z=-0.25$, $-0.5$, $-0.75$, $-1.0$, and $-1.25$. The point vortices
    all have a circulation of $\pi$, and are placed at $(1,0,0)$, $(-0.5,-0.7,0)$,
    and $(-0.5, \sqrt{3}/2, 0)$. The
    tracers are tracked for $1000$ non-dimensional time units. The Rossby
    number is $0.3$.}
    \label{fig:perturbed-triangle-varz}
\end{figure*}

The equilateral triangle configuration described earlier is a special case of a
larger class of stationary states. In general, regular polygons with identical
point vortices at each vertex rotate in place. Vortices that rotate with one
another in an approximate solid body configuration around a common center have
been discovered on other planets and have been studied by
\cite{siegelman2022b,siegelman2022a}.
Vortex polygons in two dimensions are studied in detail in \cite{aref2002}.

With identical point vortices, we can show that the distances between any
individual pair is conserved under time evolution.
\begin{multline}
  \frac{d}{dt}\rho_{il}^2
  = \frac{\I\Gamma (z_l^* - z_i^*)}{2\pi}\sum_{j=1}^N \frac{z_i - z_j}{\rho_{ij}^3} \left(1 + \frac{\rossby\Gamma}{2\pi\rho_{ij}^3}\right) \\
  + \frac{\I\rossby\Gamma^2(z_l^* - z_i^*)}{4\pi^2}\sum_{j=1}^N\sum_{k=j+1}^N\frac{2z_i - z_j - z_k}{\rho_{ij}^3\rho_{ik}^3} \\
  - \frac{\I\Gamma (z_l^* - z_i^*)}{2\pi}\sum_{j=1}^N \frac{z_l - z_j}{\rho_{lj}^3} \left(1 + \frac{\rossby\Gamma}{2\pi\rho_{lj}^3}\right) \\
  - \frac{\I\rossby\Gamma^2(z_l^* - z_i^*)}{4\pi^2}\sum_{j=1}^N\sum_{k=j+1}^N\frac{2z_l - z_j - z_k}{\rho_{lj}^3\rho_{lk}^3} \\
  + c.c..
\end{multline}
As before, the summations omit terms with $j=i$ and $k=i$.
By rewriting the distances in exponential form, we obtain
\begin{multline}
  \label{eq:polygon_rho_time_evo_sin}
  \frac{d}{dt}\rho_{il}^2
  = \frac{\Gamma\rho_{il}}{\pi}\sum_{j=1}^N\left\{\frac{\sin(\phi_{ij} - \phi_{il})}{\rho_{ij}^2} + \frac{\sin(\phi_{jl}-\phi_{il})}{\rho_{lj}^2}\right\} \\
  + \frac{\rossby\Gamma^2\rho_{il}}{2\pi^2}\sum_{j=1}^N\left\{\frac{\sin(\phi_{ij}-\phi_{il})}{\rho_{ij}^5} + \frac{\sin(\phi_{jl}-\phi_{il})}{\rho_{lj}^5}\right\} \\
  + \frac{\rossby\Gamma^2\rho_{il}}{2\pi^2}\sum_{j=1}^N\sum_{k=j+1}^N\left\{\frac{\sin(\phi_{ij}-\phi_{il})}{\rho_{ij}^2\rho_{ik}^2}+\frac{\sin(\phi_{jl}-\phi_{il})}{\rho_{lj}^2\rho_{lk}^3}\right\} \\
  + \frac{\rossby\Gamma^2\rho_{il}}{2\pi^2}\sum_{j=1}^N\sum_{k=j+1}^N\left\{\frac{\sin(\phi_{ik}-\phi_{il})}{\rho_{ij}^3\rho_{ik}^2} + \frac{\sin(\phi_{kl} - \phi_{il})}{\rho_{lj}^3\rho_{lk}^2}\right\}.
\end{multline}
The angle $\phi_{ij}-\phi_{il}$ is the angle $\angle ijl \equiv \alpha_j$ and
the angle $\phi_{jl}-\phi_{il}$ is the angle $-\angle ilj \equiv -\beta_j$.
Taking advantage of the mirror symmetry of regular polygons, we can rewrite
$\sum_{i=1}^N g_i = \frac12 \sum_{i=1}^N (g_i + g_{N-i})$. The first summation
term of \eqref{eq:polygon_rho_time_evo_sin} becomes
\begin{equation}
  \sum_{j=1}^N\left\{\frac{\sin\alpha_j}{\rho_{ij}^2}-\frac{\sin\beta_j}{\rho_{lj}^2}
  +\frac{\sin\alpha_{N-j}}{\rho_{i,N-j}^2}-\frac{\sin\beta_{N-j}}{\rho_{l,N-j}^2}
  \right\}.
\end{equation}
Because of mirror symmetry, $\alpha_{N-j}=\beta_j$, $\beta_{N-j}=\alpha_j$,
$\rho_{i,N-j}=\rho_{lj}$, and $\rho_{l,N-j}=\rho_{ij}$. This term is therefore
zero. Using these same steps, we can show that the other three summation terms
are also zero. This shows that a regular polygon with identical point vortices
at its vertices will maintain its size and shape in SQG+ under time evolution.

Because of symmetry, there will no translational motion. The rotation
rate of the regular polygon can be easily computed. Starting with the time
evolution of a point vortex \eqref{eq:z_time_evo}, we obtain
\begin{multline}
  \label{eq:omega_polygon_sqg_plus}
  \omega = \frac{\Gamma}{8\pi\rho^3}\sum_{j=1}^{N-1}\csc\frac{\pi j}{N}
  + \frac{\rossby\Gamma^2}{128\pi^2\rho^6}\sum_{j=1}^{N-1}\csc^4\frac{\pi j}{N} \\
  + \frac{\rossby\Gamma^2}{128\pi^2\rho^6}\sum_{j=1}^{N-1}\sum_{k=j+1}^{N-1}\left\{
    \csc\frac{\pi j}{N}\csc^3\frac{\pi k}{N} + \csc^3\frac{\pi j}{N}\csc\frac{\pi k}{N}
    \right\},
\end{multline}
where $\rho$ is the distance between each point vortex with the centre of the
polygon. This equation looks superficially different from (35) in
Aref 2002~\cite{aref2002}, its analogue in two-dimension. Generally, when
generalizing equations from a two-dimensional system to $\alpha$-systems, the
equation of motion is composed of terms that are power laws of distance
with different scaling exponents determined by $\alpha$.
For instance, power law relationships with a scaling exponent of $-2$ in
two-dimensions generally have a scaling exponent of $-3$ in SQG. This difference
in the scaling exponent is what led to the superficial difference between
\eqref{eq:omega_polygon_sqg_plus} and (35) of Aref 2002, where $N-1=\sum_i^{N-1}
1$ is replaced by $\sum_i^{N-1} \csc \pi j/N$.

The symmetry is preserved if we further place a point vortex of arbitrary
circulation at the centre of the polygon. Following the same steps as before
shows that distances between point vortices stay constant even in the presence of
the extra point vortex. The rotation rate of the polygon becomes
\begin{multline}
  \omega =
  \frac{\Gamma_c}{2\pi\rho^3}\left(1 + \frac{\rossby\Gamma_c}{2\pi\rho^3}\right) \\
  + \frac{\rossby\Gamma\Gamma_c}{16\pi^2\rho^6}\sum_{j=1}^{N-1}\csc\frac{\pi j}{N}
  + \frac{\rossby\Gamma\Gamma_c}{32\pi^2\rho^6}\sum_{j=1}^{N-1}\csc^3\frac{\pi j}{N} \\
  + \frac{\Gamma}{8\pi\rho^3}\sum_{j=1}^{N-1}\csc\frac{\pi j}{N}
  + \frac{\rossby\Gamma^2}{128\pi^2\rho^6}\sum_{j=1}^{N-1}\csc^4\frac{\pi j}{N} \\
  + \frac{\rossby\Gamma^2}{128\pi^2\rho^6}\sum_{j=1}^{N-1}\sum_{k=j+1}^{N-1}\left\{
    \csc\frac{\pi j}{N}\csc^3\frac{\pi k}{N} + \csc^3\frac{\pi j}{N}\csc\frac{\pi k}{N}
    \right\},
\end{multline}
where $\Gamma_c$ denotes the circulation of the central point vortex.

Plots of select passive tracer trajectories advected by point vortices arranged
in a regular pentagon are shown in Figure~\ref{fig:pentagon-corotate-ro0}. With a
point vortex at the centre of a regular polygon composed of identical point
vortices, if the central point vortex has the opposite sign as that of other
vortices, the system is not numerically stable in both SQG and SQG+.

It is worth noting that while regular point vortex polygons rotate are able to
rotate in place like a rigid object, these configurations are not robust under
perturbations.
It is well known that regular polygon arrangements of Euler 2D
point vortices and finite vortex patches are stable if $N<7$, neutrally stable
if $N=7$, and unstable if $N > 7$~\cite{havelock1931,dhanak1992}. While a
stability analysis of SQG+ point vortex polygons is beyond the scope of this
paper, our calculations show that SQG+ point vortex polygons with $7$ or more
sides are numerically unstable. The centre of mass of passive tracers advected
by a perturbed triangle is shown in Figure~\ref{fig:perturbed-triangle-varz}.
With increasing depth, the trajectory of the centre of mass becomes smoother.

\section{Vertical motion of passive tracers in SQG+}\label{sec:vertical_motion}


\begin{figure*}[t]
  \includegraphics[width=\textwidth]{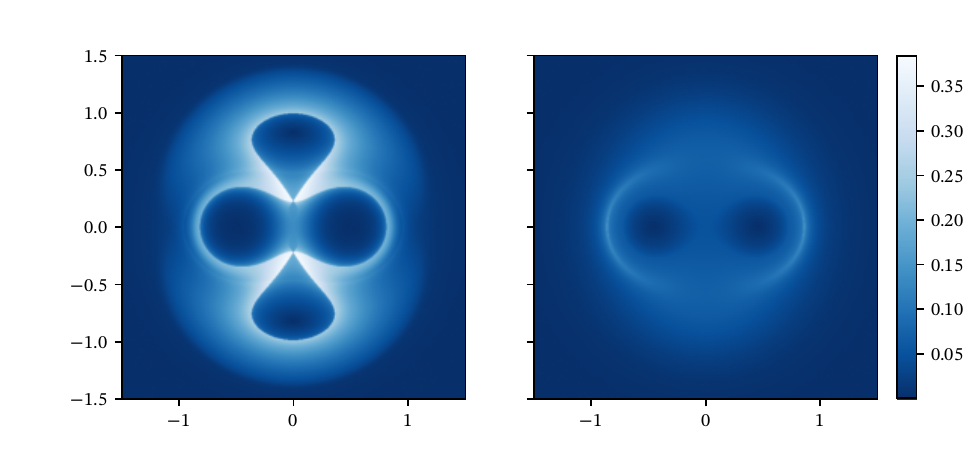}
  \caption{The maximum extent of vertical movement, or vertical excursion, of
  the passive tracers advected by two point vortices with the same strength. The
  figure on the left corresponds to the extent of vertical movement of passive
  tracers at $z=-0.25$, and the figure on the right corresponds to $z=-0.5$. The
  Rossby number is $0.3$.
  }
  \label{fig:two-pv-same-circulation-vertical-excursion}
\end{figure*}

\begin{figure*}[t]
  \includegraphics[width=\textwidth]{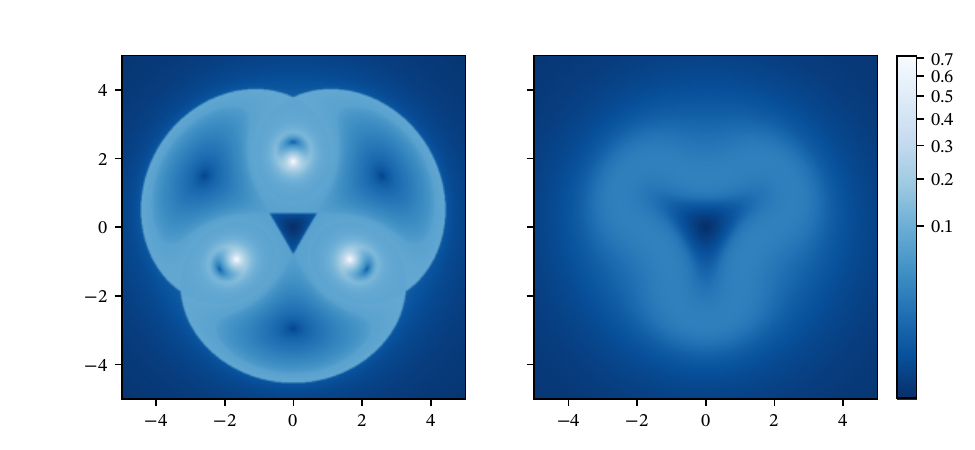}
  \caption{Vertical excursion of the passive tracers advected by three point
  vortices with strength of $\pi$. From left to right, the passive tracers are
  at depth $z=-1$ and $-2$. To better highlight the structures, a non-linear
  colour scale is used.}
  \label{fig:equilateral_triangle_vertical_excursion}
\end{figure*}

\begin{figure*}
  \begin{center}
    \includegraphics[width=\textwidth]{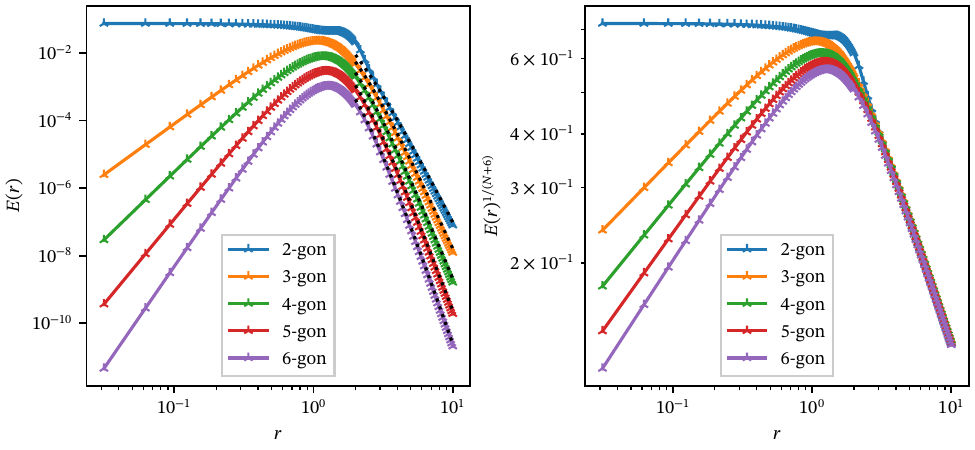}
  \end{center}
  \caption{The first plot is not scaled. The second plot is the collapsed curves.
  The black dashed curves in the first plot are given by
  \eqref{eq:vertical_excursion_r_approx}. For both plots, $\rho=1$, $\Gamma=1$,
  $z=-1.2$.}
  \label{fig:azimuthal_average_excursion}
\end{figure*}

In SQG, passive tracers have no vertical movement. A passive tracer that began
its trajectory at $z_0$ will stay in that horizontal plane. In SQG+, the
vertical velocity is given by \eqref{eq:w}. Following the approach of
Weiss~\cite{weiss2022}, we quantify the vertical movements through vertical
excursion. Vertical excursion is defined as the vertical envelope that a passive
tracer stays within throughout its trajectory. In other words, if a passive
tracer's trajectory takes it to a minimum depth $D_{\text{min}}$ and a maximum
depth $D_{\text{max}}$, its vertical excursion is
\begin{equation}
  E = D_{\text{max}} - D_{\text{min}}.
\end{equation}
The vertical excursion of passive tracers advected by two identical point
vortices is shown in Fig.~\ref{fig:two-pv-same-circulation-vertical-excursion},
and another plot for an equilateral triangle is shown in
Figure~\ref{fig:equilateral_triangle_vertical_excursion}. In both sets of plots,
the vertical movement of the passive tracers in the left panel are largely
confined to regions close to the vortices. The vertical excursion experiences a
visible drop off further away from the point vortices. This drop off with
radial distance can be quantified. In
Figure~\ref{fig:azimuthal_average_excursion}, we take azimuthal averages of the
vertical excursion and plot the results against the radial distance from the
centre. With the radii of all the vortex polygon configurations set to $1$, we
see that the vertical excursion increases from the centre of the plane to a peak
at $=1$, then enters a drop off that follows the power law. It is apparent that
the power law that governs the radial decrease
in vertical movement scales with the number of point vortices present.

\begin{figure*}
  \begin{center}
    \includegraphics[width=\textwidth]{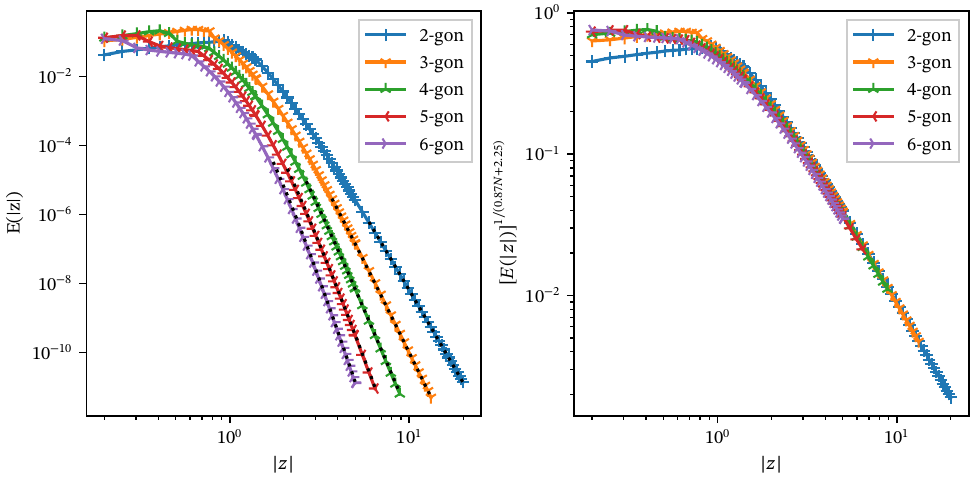}
    \end{center}
    \caption{Numerical solutions of the vertical excursion. The vertical
    excursions were computed at a fixed $r=1$ and averaged over a ring. $\Gamma
    = 1$. The first plot is the unscaled vertical excursion. The second plot is
    the collapsed curves. The black dashed curves in the first plot are given by
    \eqref{eq:vertical_excursion_z_approx}.}
    \label{fig:vertical-excursion}
\end{figure*}

Comparing the right panel to the left panel of both
Figures~\ref{fig:two-pv-same-circulation-vertical-excursion} and
\ref{fig:equilateral_triangle_vertical_excursion}, a decrease in
vertical excursion can be seen when the depth is increased. This decrease can be
quantified. In Figure~\ref{fig:vertical-excursion}, we take the azimuthal
average of the vertical excursion of passive tracers at a radial distance $r=1$
and plot the results against $z$. We again see a power law decrease of the
vertical excursion with increasing depth. This power law also scales with
the number of point vortices present.

\begin{table}
  \begin{center}
    \begin{tabular}{lccccc}
      \toprule
      N & 2 & 3 & 4 & 5 & 6 \\
      \midrule
      A & $3$ & $135/8$ & $102/5$ & $7578/64$ & $31185/128$ \\
      \bottomrule
    \end{tabular}
  \end{center}
  \caption{Expansion coefficients of the far-field vertical velocity. There is
  no apparent pattern between them.}
  \label{tbl:a}
\end{table}

While the complexity of \eqref{eqs:velocity_equations} and \eqref{eq:w} makes it
difficult to directly explain how vertical excursion changes with increasing
depth or horizontal distance from the vortices, we can study the far-field
vertical excursion through asymptotic analysis. Far away from the point vortex
polygon where $r^2 + z^2 \gg \rho^2$, where $r$
and $z$ are the radial and vertical positions of the passive tracer and $\rho$
is the radius of the $N$-sided regular polygon, we can expand $u$, $v$ and $w$
in orders of $\rho / \sqrt{r^2 + z^2}$ to get a clear picture of the behavior
far from the origin. We first expand the horizontal velocity up to the leading
order. Given that the SQG+ correction solely consists of higher order terms, we
omit the SQG+ corrections from these derivations.
\begin{align}
  \begin{split}
    \bfvec{u} &= \sum_j^N \frac{\bfvec{\hat{k}}\times\Gamma (\bfvec{x} - \bfvec{x}_j)}{2\pi\left|(x-x_j)^2 + (y-y_j)^2 + z^2\right|^3} \\
    &= \begin{multlined}[t]
      \bfvec{\hat{k}}\times \frac{\Gamma}{2\pi(r^2 + z^2)^{3/2}} \\
    \times \sum_j^N (\bfvec{x}-\bfvec{x}_j)\left(1 - \frac32 \frac{\rho^2 - 2r\rho\cos\left(\theta - 2\pi j/N\right)}{r^2 + z^2} + \ldots\right)
    \end{multlined} \\
    &\approx \bfvec{\hat{k}}\times \frac{N\Gamma\bfvec{x}}{2\pi(r^2 + z^2)^{3/2}}.
  \end{split}
\end{align}
It is clear that the horizontal velocity of the passive tracers far away from
the point vortices has no radial components. A passive tracer follows circular
orbits around the origin with angular velocity
\begin{equation}
  \omega_f \approx \frac{N\Gamma}{2\pi(r^2 + z^2)^{3/2}}.
\end{equation}
Far away from the $N$-sided vortex polygon, passive tracers revolve around the
polygon as if the polygon is one SQG point vortex with a circulation of
$N\Gamma$.

\begin{figure*}
  \begin{center}
    \includegraphics[width=\textwidth]{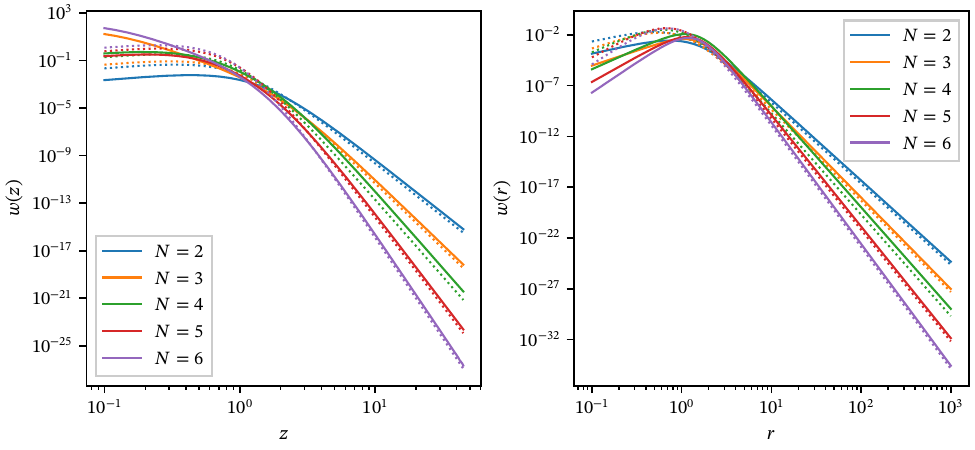}
  \end{center}
  \caption{Exact SQG+ solution of $w$ vs.~its first order approximation. The
  solid lines are the exact SQG+ solutions while the dashed lines are the
  leading order approximations. $\Gamma = 1$, $\theta = 2$ and $\rho = 1$ for
  both plots. $r = 1$ for the first plot and $z = 1$ for the second plot.}
  \label{fig:w-trial}
\end{figure*}

The vertical velocity is determined solely by SQG+ terms, given in \eqref{eq:w}.
Changing to cylindrical coordinates and
expanding the equation in $\rho / \sqrt{r^2 + z^2}$, we obtain
\begin{equation}\label{eq:w_approx}
  w \approx \frac{A \Gamma^2 r^N \rho^N z \sin N \theta}{\pi^2 (r^2 + z ^2)^{N + 3}},
\end{equation}
where $\theta$ is the azimuthal position of the passive tracer in polar
coordinates. Due to the complexity of the expansion, it is quite difficult to
find a general formula for the scaling coefficient $A$. Instead, a list of
values for $A$ for the first few regular polygons is given in Table~\ref{tbl:a}.
A visual comparison of the full SQG+ vertical velocity \eqref{eq:w} and
the far-field approximation \eqref{eq:w_approx} is shown in
Figure~\ref{fig:w-trial}. The approximation shows good agreement with the SQG+
solution. Writing $\theta = \omega_f t + \theta_0$, we obtain an ordinary
differential equation that describes the trajectory of a passive tracer far away
from the point vortices,
\begin{equation}\label{eq:z}
  \frac{\diff z}{\diff t} = \frac{A \Gamma^2 r^N \rho^N z}{\pi^2 (r^2 + z ^2)^{N + 3}}
  \sin \left(N \left(\frac{N\Gamma}{2\pi(r^2 + z^2)^{3/2}}t + \theta_0\right)\right).
\end{equation}
To find the vertical excursion analytically, in theory, one only needs to
integrate this equation. While this equation is not separable, we can  see how $N$ enters the
approximate scaling laws given above. For large depth, by
performing a curve-fit on the first plot in Fig.~\ref{fig:vertical-excursion},
we are able to obtain
\begin{equation}\label{eq:vertical_excursion_z_approx}
    E(z) \approx \frac{(8.27 - 0.94 N) z}{(1 + z^2)^ {0.87 N + 3.25}},
\end{equation}
for sufficiently small values of $N$. The approximated curves are shown as dashed
curves in the same plot. With this expression, we are able to collapse the curves
onto one curve in the second plot. Likewise, for large $r$, we obtain from the data
in Fig.~\ref{fig:azimuthal_average_excursion}
\begin{equation}\label{eq:vertical_excursion_r_approx}
    E(r) \approx \frac{((10.34-N)(N-2)+10.5)r^N}{(1.44 + r^2)^{N+3}}.
\end{equation}

\begin{figure*}
    \begin{center}
        \includegraphics[width=\textwidth]{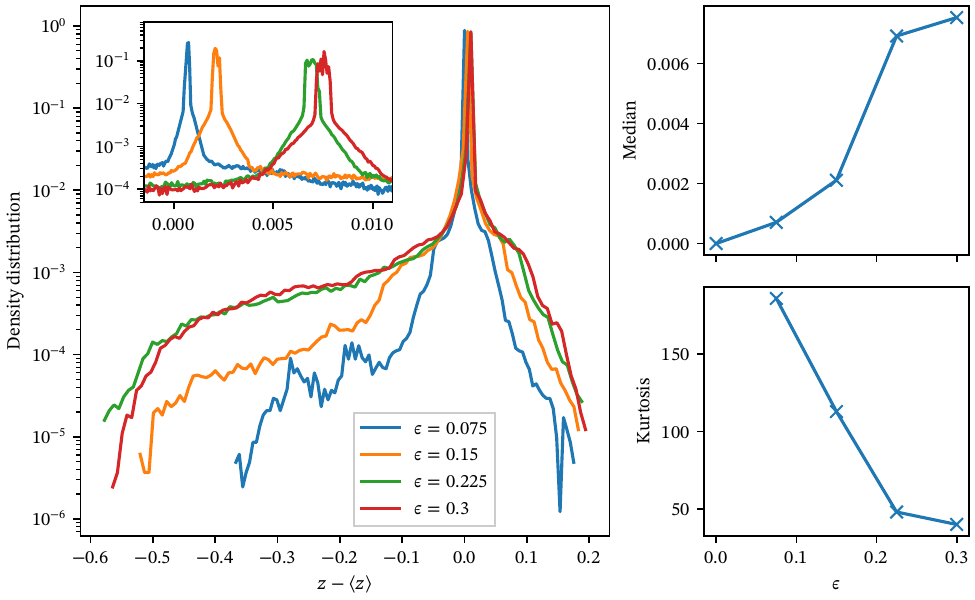}
    \end{center}
    \caption{The statistics of passive tracers advected by a perturbed triangle.
    The setup is identical to that of Figure~\ref{fig:perturbed-triangle-varz}.
    The probability distribution function is averaged over the final $20$
    non-dimensionalised time units. The inset shows a magnified view of the
    distributions near the peaks.
    }
    \label{fig:perturbed_triangle_vertical_pdf_v_epsilon}
\end{figure*}

Without any of the aforementioned symmetries, the mathematics become more opaque.
However, we are still able to describe the vertical motion through statistics.
Using the same configuration as Fig.~\ref{fig:perturbed-triangle-varz}, we analysed
the distribution of the vertical distance of the passive tracer from the centre of
mass in Fig.~\ref{fig:perturbed_triangle_vertical_pdf_v_epsilon}.
In concrete terms, this quantity is given by
\begin{equation}
    Dz_i(t) = z_i(t) - \frac{1}{N} \sum_j z_i(t),
\end{equation}
where $N$ is the number of passive tracers present, and $z_i$ is the vertical
position of the $i$-th passive tracer. The first plot is the density
plot of $Dz_i$. The second plot is a plot of the peaks of this density function
as a function of the Rossby number $\epsilon$. The third plot shows the
kurtoses of these density functions. The density distribution of $\epsilon=0$ is not
shown, since there is no vertical movement, and the density function is a Dirac delta
function centred at $z=0$. With increasing Rossby number, the peak of the probability
distribution function moves further away from the centre of mass toward positive $z$.
This is clearly visible in the inset diagram, and the upper right plot shows how
the median shifts with increasing Rossby number. The sharp peaks of the distributions
get progressively less sharp with increasing Rossby number, as shown by decreasing
kurtosis. Physically, this means the cloud of passive tracers gets thicker with
increasing Rossby number. The vast majority of the passive tracers do not stray very far from the starting vertical position, but a cloud of tracers extend further away from this position with increasing Rossby number. This shift is not symmetric, however. For this cloud of passive tracers, there is a clear preference for downward expansion because the surface above limits the tracers' upward trajectory.

\section{Discussion}\label{sec:conlusions}

In this study, we have found that unlike the SQG case, total linear and angular
momenta are not conserved for SQG+ point vortices.  While two unequal point
vortices produce aperiodic passive tracer trajectories that are not divergence-free at the surface, ruling out the existence of a streamfunction in SQG+ for
this configuration, the existence of a Hamiltonian cannot be dismissed.
There is a large literature on Hamiltonian fluid dynamics~\cite{morrison1998}.
Since SQG+ dynamics is built upon an asymptotic expansion and balance, the
Hamiltonian structure in the primitive equations should be
preserved~\cite{weiss2022}. We leave the exploration of Hamiltonian structures
of SQG+ point vortices for future work.

Beyond the dynamics driven by two point vortices, we derived the equations
of motion for vortex polygons and collinear vortices. We found that SQG+ vortex
polygons and collinear vortices behave in very similar ways as their Euler 2D
counterpart. While a full stability analysis of such configurations is beyond
the scope of this study, we have demonstrated the robustness of SQG and SQG+
vortex polygons of $N \leq 6$ under perturbations. This is analogous to the
Havelock results. We believe a full stability analysis for SQG+ point vortex
polygons and vortex patch polygons will be a natural next step to explore, which
we will leave to a future study.

We examined passive tracer trajectories along with the effect of the order
Rossby correction on vertical motion. Vertical motion is what truly sets SQG+
apart from Euler 2D and SQG flows. We used vertical excursion as a proxy to
quantify how much passive tracers are being moved in the vertical
direction. In addition, we derived the far-field vertical behaviour of
passive tracers for certain special cases. We obtained the asymptotic expansion
of the velocity components far away from the point vortices for the special case
of vortex polygons and found scaling laws for vertical excursion in the
limit of large $r$ and $z$ through numerical experiments. The
magnitude of vertical velocity and by association, vertical movement, decreases
more quickly with an increasing number of point vortices in the vortex polygon.
This is surprising, since far away from the surface, the horizontal velocity
decays according to $z^{-3}$. Given this result, one would expect the vertical
velocity to scale similarly with a power law that equally applies to all
polygonal configurations. However, when we expand the vertical velocity equation, the lowest order terms that don’t vanish in the expansion can all be written in the form
\begin{equation}
    \frac{B r^N z \sin N \theta}{(r^2 + z^2)^{N + 3}},
\end{equation}
so the power law scales with the number of point vortices. This is a result of the rotational symmetry of the polygon.
One can expect a similar scaling behaviour for the SQG+ analogue of the finite vortex patch polygons studied by Dhanak~\cite{dhanak1992}.
One might even find this behaviour in more general situations where there is a similar symmetry.
This is an open question worth
exploring.

\section*{Acknowledgements}

The authors would like to acknowledge the inspiring discussions with
W.~R.~Young.

\section*{CRediT}
\textbf{Mac Lee}: Software, Formal analysis, Investigation, Writing - Original
Draft \textbf{Stefan Llewellyn Smith}: Conceptualization, Methodology,
Validation, Writing - Review \& Editing, Supervision, Funding acquisition

\section*{Funding}

This research was partially funded by an ONR NISEC award.

\section*{Declaration of competing interests}

The authors report no conflict of interest.

\section*{Code availability}

Code to generate the data used in this paper is available at
\url{https://github.com/macthecadillac/point-vortex}.

\appendix
\section{SQG+ collinear criteria}\label{sec:collinear-criteria}

\begingroup
  \allowdisplaybreaks
  \begin{align}
      A &= \frac{\rossby'}{2\pi}(\Gamma_2^2+\Gamma_3^2) \\
      B &= \frac{\rossby'}{2\pi}(5\Gamma_2^2+6\Gamma_3^2) \\
      C &= \frac{\rossby'}{2\pi}(\Gamma_1\Gamma_2-\Gamma_1\Gamma_3+10\Gamma_2^2+15\Gamma_3^2) \\
      D &= \Gamma_2+\Gamma_3+\frac{\rossby'}{2\pi}\left(4\Gamma_1\Gamma_2-5\Gamma_1\Gamma_3+10\Gamma_2^2+20\Gamma_3^2\right) \\
      E &= 2\Gamma_2+3\Gamma_3+\frac{\rossby'}{2\pi}\left(5\Gamma_1\Gamma_2-9\Gamma_1\Gamma_3+5\Gamma_2^2+15\Gamma_3^2\right) \\
      F &= \Gamma_2+3\Gamma_3+\frac{\rossby'}{2\pi}\left(2\Gamma_1\Gamma_2-5\Gamma_1\Gamma_3+\Gamma_2^2+6\Gamma_3^2\right) \\
      G &= -\left\{3\Gamma_1+\Gamma_2+\frac{\rossby'}{2\pi}\left(6\Gamma_1^2-5\Gamma_1\Gamma_3+\Gamma_2^2+2\Gamma_2\Gamma_3\right)\right\} \\
      H &= -\left\{3\Gamma_1+2\Gamma_2+\frac{\rossby'}{2\pi}\left(15\Gamma_1^2-9\Gamma_1\Gamma_3+5\Gamma_2^2+5\Gamma_2\Gamma_3\right)\right\} \\
      I &= -\left\{\Gamma_1+\Gamma_2+\frac{\rossby'}{2\pi}\left(20\Gamma_1^2-5\Gamma_1\Gamma_3+10\Gamma_2^2+4\Gamma_2\Gamma_3\right)\right\} \\
      J &= -\frac{\rossby'}{2\pi}\left(15\Gamma_1^2-\Gamma_1\Gamma_3+10\Gamma_2^2+\Gamma_2\Gamma_3\right) \\
      K &= -\frac{\rossby'}{2\pi}(6\Gamma_1^2+5\Gamma_2^2) \\
      L &= -\frac{\rossby'}{2\pi}(\Gamma_1^2+\Gamma_2^2),
  \end{align}%
\endgroup
where $\rossby'=\rossby/\rho_{13}^3$.

\bibliographystyle{elsarticle-num}
\bibliography{ref_els}
\end{document}